\documentclass[12pt]{article}
\usepackage{cite}

\newcommand{\be}{\begin{equation}}
\newcommand{\ee}{\end{equation}}
\newcommand{\bea}{\begin{eqnarray}}
\newcommand{\eea}{\end{eqnarray}}
\def \ci{\cite}

\newcommand{\nn}{\nonumber}
\newcommand{\p}[1]{(\ref{#1})}
\newcommand{\bt}[1]{{\bar t}}

\def \bc  {{\hat  {\rm c}}}

\usepackage{epsfig,color,latexsym}

\def \sql {{\sqrt{\l}}\ }
\def \del{\partial}
\def \a {\alpha}

\def\s{\sigma}

\def\ov{\over}
\def\la{\label}

\def\J{{\cal J}}

\def\l{\lambda}
\def\eps{\epsilon}

\def \adss{$AdS_5 \times S^5$\ } 

\def \r { \rho}
\def \sql {\sqrt{\lambda} }

\def \p {\phi}
\def \vp {\varphi}

\def \ov {\over}
\def \s{\sigma}

\def \ta{\tau}
\def \sh {\sinh}

\def \la{\label}

\def \k {\kappa}
\def\foot{\footnote}

\def \J {{\cal J}}
\def \L {\Lambda}

\def\a{\alpha }
\def\ka{\kappa }
\def\s{\sigma }
\def\J{{\cal J}}

\def\S{{\cal S}}
\def\ch{\cosh}
\def\th{\tanh}
\def\sh{\sinh}
\def\p{\partial}
\def \ov {\over}
\def \del {\p}
\def \foot{footnote}

\newcommand{\rf}[1]{(\ref{#1})}
\renewcommand{\theequation}{\thesection.\arabic{equation}}
\renewcommand{\thefootnote}{\fnsymbol{footnote}}

\def\appendix#1{
  \addtocounter{section}{1}
  \setcounter{equation}{0}
  \renewcommand{\thesection}{\Alph{section}}
  \section*{Appendix \thesection\protect\indent \parbox[t]{11.15cm}
  {#1} }
  \addcontentsline{toc}{section}{Appendix \thesection\ \ \ #1}
  }

\def\be{\begin{equation}}
\def\ee{\end{equation}}

\def \ci {\cite}

\def \foot {\footnote}
\def \bi{\bibitem}

\def \td {\tilde}

\def \ci{\cite}

\def \xe  {x_{0e}}
 
\def\S{{\cal S} }
\def \bra {\langle}
\def   \ket  {\rangle}

\def \rX {{\rm X}}

\def \rY{{\rm Y}}

\def \rx {{\rm x}}
\def \ov {\over}
 \def \D {\Delta}
\def \te {\tau_e} 
\def \be  {\bea}
\def \ee {\eea} \def \ep {\epsilon}\def \G {\Gamma}
\def \calJ {{\cal J}}
\def \ax {\varepsilon}
\def \rc {{\rm c}}
\def \bc {{\bar {\rm c}}}
\def \bc  {{\hat  {\rm c}}}

\begin{document}


\null\vskip-24pt \hfill ICCUB-10-199
\vskip-1pt
\hfill
Imperial-TP-AT-2010-04 
\vskip-1pt

\vskip.5truecm

\vskip0.2truecm
\begin{center}
\vskip 0.2truecm {\Large\bf
Large spin expansion 
\vskip 0.2truecm
of semiclassical 3-point correlators in $AdS_5\times S^5$
}
\vskip 1.5truecm
J.G. Russo$^{a,b}$
and A.A. Tseytlin$^{c}$\footnote{Also at   Lebedev
 Institute, Moscow.}\\
\vskip 0.4truecm
\vskip .2truecm
$^{a}$  
 Instituci\'o Catalana de Recerca i Estudis Avan\c cats (ICREA)\\
Pg. Lluis Companys, 23, 08010 Barcelona, Spain.\\
\smallskip
$^{b}$  Institute of Cosmos Sciences and ECM, Facultat de F{\'\i}sica,\\ Universitat de Barcelona,
Diagonal 647,  08028 Barcelona, Spain.\\
\smallskip
$^{c}$ Blackett Laboratory, Imperial College,
London SW7 2AZ, U.K.

\end{center}
\vskip 0.5truecm
\vskip 0.2truecm \noindent\centerline{\bf Abstract}
\vskip .2truecm 



3-point correlators  in $AdS_5\times S^5$
string theory  in which two states are ``heavy'' (have   large  quantum numbers) and  the third  is ``light''
(here chosen as chiral primary scalar)  can be computed semiclassically 
in terms of 
the  ``light" vertex operator evaluated on the classical 
 string solution sourced by the two ``heavy'' operators.
 We observe that in the case when  the ``heavy'' operators  represent  BPS states 
 there is an ambiguity in the computation  depending 
 on  whether the mass shell (or marginality) condition is imposed  before or
  after  integration over the world sheet.  We show that  this ambiguity
 is resolved in a universal way by defining the  BPS  correlator as 
 a limit  of the one with  non-BPS  ``heavy'' states.   
  We consider several  examples  with ``heavy'' states  represented by folded 
  or circular spinning strings in $AdS_5\times S^5$   that 
  admit a point-like BMN-type limit 
  when one $S^5$ spin $J$ is much larger than the others. 
  Remarkably, in  all of these cases the large  $J$  expansion of  the 3-point correlator 
   has the same structure  as expected in  perturbative (tree-level and one-loop)
  dual  gauge theory. We conjecture that, like the leading  chiral
   primary correlator term,  the  coefficients of the first few  subleading 
     terms  are also protected, i.e. should be the same 
     at strong and weak coupling.


\newpage

\renewcommand{\thefootnote}{\arabic{footnote}}
\setcounter{footnote}{0}

\def \ed {\end{document}}

\section{Introduction}

A complete proof  of (planar) AdS/CFT duality requires matching not only 2-point
 but also 3-point correlation functions of primary operators 
 as they determine the  structure of a  CFT. 
During the last few years  an impressive progress was achieved in 
understanding the  integrability-based   equations determining   dimensions of general 
non-BPS operators (and thus their 2-point functions) 
 at any  value of  gauge coupling or string tension (see, e.g., \ci{revi}).
 However, 
the properties of tree-level 
string and planar gauge theory 3-point functions remain largely an 
unexplored area.

Recently, there was a renewed  interest in 2-point and  3-point correlators 
in $AdS_5 \times S^5$ string theory in the semiclassical limit 
of large quantum numbers   (see, e.g., \ci{b,jan,bt1,z,cost,rt,ry}). 
 In the case  when two of the three string vertex operators  
 represent  ``heavy''  states that 
  carry large charges  while the third  corresponds to   a  ``light''  BPS 
   state
   with   
  fixed charge it was
 suggested in \ci{z,cost} that such  3-point correlator   can
  be computed by evaluating the ``light''  operator on the classical solution
  ``sourced'' \ci{tse03}  by the ``heavy'' operators. 
   This  observation 
  was further generalized  and extended    to the case when the  ``light'' 
  operator may     represent a non-BPS  string state  in \ci{rt}.

  Such  semiclassical method of computing 3-point functions involving 
two ``heavy'' and one ``light'' state 
can be explicitly tested in  string theory in flat space-time \cite{Chialva:2004xm}. 
Such correlator describes, e.g., 
a decay of  rotating massive string by radiation emission.
For example, for a specific state representing a 
circular string rotating in several planes, one can independently 
carry out  an  exact (tree-level)   quantum string calculation
 in terms of creation/annihilation operators, producing  a  rather non-trivial 
expression in terms of  an infinite sum. In the  limit of large
 mass of the ``heavy'' state, in which 
 the semiclassical approach should be  reliable, the semiclassical
 calculation reproduces {\it exactly}  the result of the quantum 
 string calculation.

  \

Here we   shall follow  closely the  approach of \ci{rt}. 
Our aim  will be to analyse the large angular momentum expansion 
 of the 
semiclassical expressions for correlation 
functions of two ``heavy'' non-BPS operators  and one
``light'' chiral primary operator.

 It was noticed in \ci{z} that the  BMN-type limit   ($J_1 \gg J_2$)  
of   such  3-point function
involving a ``heavy'' state   corresponding to the folded  spinning ($J_1,J_2)$
 string on $S^5$   
 containes an extra term as compared to the correlator of the 
 three BPS  (chiral primary) operators computed directly.  
 This extra  term  may be interpreted as  arising 
 from  a ``$ 0 \ov 0$ -type''  ambiguity.   Here we shall  clarify the universal nature
 of this ``anomalous'' contribution showing that it always 
 appears for 
 generic  string states  that admit a  point-like BMN limit. 
 We shall see that  to have a smooth BPS limit of a non-BPS correlator one 
 may formally 
 adjust  the normalization of the  ``light'' chiral primary operator.
 

We will also  study  the subleading terms in the large  spin  expansion 
of the  semiclassical 3-point correlator. 
In general, the  semiclassical string expansion is based  on taking 
string tension  large, i.e. $\sql \gg 1$, while   fixing the 
classical spin parameters $ \J_i = { J_i \ov \sql}$. 
While this limit  is, of course,  different from the large $J$ limit in
 perturbative gauge theory  (where  one first expands in $\l \ll 1$ and then takes 
 $J_i$ large) some leading terms in the  string and gauge theory observables expanded 
 at large $J$  may happen to have the same structure. Morever, 
 the coefficients  of these terms    may happen to be protected (due to supersymmetry), 
 i.e.  may  be the same at strong and weak coupling.  
 This is, indeed,   what  happens, e.g., 
in the expansion of  the energy of a classical 
string on $S^5$ with two spins   in the limit $J_1 \gg J_2$
when the c.o.m.  orbital momentum  is large \ci{ft2}:
$E= J  + c_1 { \l \ov J}  + c_2 { \l^2 \ov J^3} + ...$,
 where $J= J_1 + J_2$  and
  $c_i$ are  functions of ${ J_2 \ov J}$
  (one  may  then further  expand  in ${ J_2 \ov J} \ll 1 $
  which would correspond to approaching the BMN limit).
Not only this expansion has the same   form as the corresponding
large $J$ expansion in the perturbative gauge theory,  but also  the first few
leading coefficients $c_i$  match exactly, i.e.  are protected against
$\l$-dependent corrections \ci{ft2,bmsz,Frolov:2003xy,bfst,bt,bes}
(for a review, see, e.g., \ci{gio}).

Since  the  one-loop  corrections to 3-point correlators in dual gauge
 theory
often have structure similar to one-loop anomalous dimensions of the
operators involved  (see, e.g., \ci{ok,okk})
one may  wonder if the leading  terms in the
large spin expansion of
the semiclassical string-theory correlators may also follow the  pattern
observed in the expansion of semiclassical string energies.
Indeed,
the coefficients in the  semiclassical string  3-point correlators
can be expanded  in a similar  way  in
powers  of $ {1\ov \J} = { \sql \ov J}$
for fixed ${ J_2\ov J}$. In addition, one may
 further  expand  in ${ J_2 \ov J} \ll 1 $, i.e. consider a near BMN limit.
As we shall demonstrate below on several examples,
the   large  $J$  expansion of  the semiclassical string  3-point correlators 
   has formally  the same structure  as would be expected in 
    gauge theory. 
  We are then led to   conjecture that, like the leading (three  chiral
   primary correlator) term,  the  coefficients of the first few  subleading 
     terms in this expansion 
     may   also protected, i.e.  may turn out to  be the same  
      on the perturbative string and the perturbative gauge theory sides.  
 It would be very interesting to check this 
 conjecture by direct computation of the corresponding 
  correlators on the weak coupling gauge theory side. 
 This should hopefully  become possible soon using the integrability-based approach 
 suggested  in  \ci{gr}.


\


The structure of this paper  is as follows.
In section 2 we review the basic setup for the semiclassical calculation of
 the correlators in $AdS_5\times S^5$ involving
two ``heavy'' and one ``light'' (chiral primary)  state and
discuss the ambiguity that arises in the case  when all three  operators are 
BPS (chiral primary) ones and suggest its resolution. 
As a test of our proposal, in section 3 we show that in the case 
when the ``heavy'' states are represented  by  
rigid strings of arbitrary shape rotating on $S^5$
the ``anomalous'' rescaling of the correlator  in the large spin 
limit can  be removed by 
adjusting  the  normalization of the ``light''   chiral primary vertex operator. 

To study  subleading terms in the large  spin expansion of the semiclassical correlator
in  section 4 we consider  few  explicit examples:  
folded string with two angular momenta extended on $S^5$, folded string extended
 on $AdS_5$ with rotation in both $AdS_5$ and $S^5$, circular strings
extended and rotating in both  $AdS_5$ and $S^5$.
We   show that the large $S^5$ spin  expansion  can be  organised as expansion 
in powers of $\lambda\ov J^2$, i.e.  has the  same structure  as expected 
in  perturbative gauge theory.


\renewcommand{\theequation}{2.\arabic{equation}}
 \setcounter{equation}{0}
 
\section{ Ambiguity  in semiclassical computation of  correlator of  three
   chiral primary  operators } 

Let us start   with revisiting the computation of the correlator of 
three  chiral primary operators with charges $ J_1 \approx 
J_2\equiv J  \gg J_3\equiv j$. Following \ci{z,rt}  the coefficient in 
the 3-point function can be found by
evaluating the vertex operator  corresponding to the ``light''  state 
with  $SO(6)$  orbital momentum  $j$
on the classical   solution ``sourced''  by the  two ``heavy'' 
BMN operators with charges $J, -J$.\foot{Below we shall assume that $J  >0,\ j > 0$.
The charge conservation requires
that the operators   have charges $J, -J-j, j $  but for 
$ J \gg j$ ones   has  $-J-j\approx -J$.}

The relevant  part of the integrated  CPO 
vertex operator \ci{bcfm,z} inserted 
at point $\rx^m =0$ at the boundary of the Poincar\'e patch of $AdS_5$
($ds^2 = z^{-2} ( dz^2 + dx^m dx_m$) 
 is \ci{rt} 
\foot{We shall  follow closely the notation in
\ci{rt}  were details may be found. Here we consider the Euclidean continuation:
$\tau_e=i \tau$. Compared to \ci{rt} here we extract the factor of  string tension
$T = { \sql \ov 2 \pi} $  from
$\rc_j$. \ $N$  here  stands for a factor of string coupling or 
rank of gauge group.}
 \bea 
&& V^{(cpo)}_\Delta (0) = \bc_j \int^\infty_{-\infty}
 d \tau_e \int^{2\pi}_0 d \sigma\
  (\rY_+)^{-\D}\  \rX_x^j \   
U^{(cpo)}(x,z, X) \ , \ \  \   \ \ \   \Delta = j \  ,   \la{kz} \\
 &&   U^{(cpo)}= { \sql \ov 2 \pi}\ \Big[
    z^{-2} (\del^a x^m \del_a   x_m   -  \del^a z   \del_a z )
     -  \del^a X_k    \del_a  X_k \Big]   \  , \ \ \ \ \ \ \ \ \ 
     \bc_j = {\sqrt j (j+1) \ov 4N} \ .  \la{koz} \eea  
Here  \bea  \la{dee}
 \rY_+ = Y_5 + Y_4      =  z+ z^{-1} x^m x_m\  ~, \ \ \ \ \ \ \ \
 \rX_x= X_1 + i X_2= \cos  \vartheta \  e^{i \vp}
 \ ,   \eea
 where  $Y_M$ and  $X_k$ are  embedding coordinates of $AdS_5$ and $S^5$ 
 and  
  $\rc_j$ is a normalization constant of the chiral primary  vertex operator 
  that will be discussed later.
 For a  ``heavy''-state classical solution  satisfying $z^2 + x^m x_m=1$ 
 (all solutions discussed below will be of that type) 
 one then finds the 
 3-point correlator coefficients  as \ci{rt}\foot{As one can show  \ci{bt2},
 in the case when  the classical solution  is ``sourced''  
 by the vertex operators inserted at the points $\rx_1$ and $\rx_2$   with 
 $\rx_1 =-\rx_2$ one  gets the same semiclassical result if one uses 
 the full CPO vertex in \ci{z} or its truncated version in \rf{koz}.
 We have  checked this explicitly for the examples considered below.}
 \be 
  C_{123} &=& 
  {\bra  V_{H} (\rx ) V_{H} (-\rx) V^{(cpo)}_\D(0) \ket 
    \ov \bra  V_{H} (\rx) V_{H}(-\rx)\ket } \nonumber \\
    &=&   \rc_j \ 
 \Big\{  \int^\infty_{-\infty} d\tau_e  \int^{2\pi}_0 
  d\s \  z^{j}\  \rX_x^j\  U^{(cpo)}(x,z, X) 
  \Big\}_{class} \ , \ \ \ \ \ \ \ \ \ \ \ \    \rc_j = 2^{-j} \ \bc_j            \ ,  \la{jpl} \ee
where we assumed that  $|\rx|=1$   and 
the r.h.s. is evaluated on the classical solution determined by the 
``heavy'' state.

The classical solution  ($t= \k \tau, \ \vp = \nu \ta, \ \k=\nu$) 
describing a point-like string  with  large orbital 
momentum  $J$  in $S^5$ is 
 a massive geodesic in  $AdS_5$, running through its center.
 Written in the Poincar\'e coordinates, 
after a euclidean continuation it is 
\bea 
&&z=\frac{1}{\cosh(\kappa \tau_e)}\,, \qquad \xe =\tanh(\kappa \tau_e)\,,
\ \ \ \qquad
x_i=0\,,  \label{ga} \\
&&   \vp = -i \nu \tau_e \ , \ \ \ \  \ \ \ \    \k^2 = \nu^2   \  , \ \ \ \ 
\ \ \ \ \ \    J = \sql\ \nu \ . 
\label{ge}  
 \eea
The radial coordinate $z$ vanishes in the limits  $\tau_e \to \pm \infty$, 
implying  that the euclidean trajectory 
reaches the boundary at the two points:  $\xe=-1,\ x_i=0$ and  $\xe=1,\ x_i=0$.
Evaluating \rf{jpl} on this solution, but without imposing 
the ``on-shell''  (or Virasoro) condition $\k^2 = \nu^2$,   we get
\foot{Let us 
 comment on the  euclidean continuation $\tau \to \tau_e$.
The original integrand has a factor $1\ov \cos \ka \tau $ which
 produces poles on the real axis at $\kappa \tau  =
 {\pi\over 2}(2n+1)$, with $n$=integer.
The Wick rotation is performed by choosing 
a closed contour that does not enclose any pole; 
for  ${\rm Re}\ \tau >0$  this is done by 
adding small semi-circles lying on ${\rm Im}\ \tau >0$ surrounding 
the poles and closing the contour by a  quarter-circle at 
infinity on the upper plane.
Similarly,  for  ${\rm Re}\ \tau <0$, the contour is closed by a  
quarter-circle lying on ${\rm Im}\ \tau <0$
and adding small semi-circles around the poles lying on
 ${\rm Im}\ \tau <0$ in such a way no pole is enclosed.
The contributions of
 the upper and lower quarter-circles  vanish because 
 $|\kappa|>\nu$ so that 
$e^{ij \nu \tau }\ov \cos^j \kappa \tau $ goes to
zero as ${\rm Im}\ \tau \to\pm \infty$. Thus one 
is left with the integral over  the imaginary axis, i.e. 
 the integral over  $\tau_e $ from $-\infty$ to $\infty $
appearing  in the above expressions below.} 
\bea 
&& C_{123} = \rc_j\sql   \int^\infty_{-\infty} d\tau_e\  {e^{ j \nu\te } \ov \cosh^j \k \te }
 ( U_1 +   U_2)  \ , \la{ki} \\
 &&
 U_1 = { 2 \k^2 \ov \cosh^2 \k \te}  \ , \ \ \ \ \ \ \ \ \ \ \ 
 U_2 =  - (\k^2  - \nu^2)  \   . \la{ju}
  \eea 
We have  not used   the  condition $\k^2 = \nu^2$  to illustrate that 
the   result  will be {\it different} if \rf{ki} is computed  using the 
two different procedures:
 \medskip
 
(I) if the on-shell condition $\k^2 = \nu^2$ is first used  in the integrand;
\smallskip

(II) if one first  does  the integral and  then  takes
 the limit $\k^2 - \nu^2 \to 0$. 

\medskip
 
 \noindent 
 The analog of the  $U_2$ term  in \rf{ju}
   will  always 
   appear  if instead  of a point-like string we consider a ``heavy''
    state represented
   by a 
 small  string spread in $S^5$:   then  the  corresponding 
 Virasoro condition will 
 be $\k^2  - \nu^2= O(\ep^2)$   where $\ep$ is the 
 string size (see below). 
 If one then takes the point-like string limit $\ep\to 0$,
 one obtains the  result which agrees with the $C_{123}$ of the
 point-like string  (\ref{ki}), (\ref{ju}) computed with the prescription (II).
 This  ``anomaly'',  i.e. the difference 
 in  the outcome  of the two prescriptions  was first 
  observed in \ci{z} on  the example of the folded  2-spin  string
 in the limit when  the spin around its  center of mass  goes to zero.
 Our  aim below   will be  to  expose the universal nature of this ``anomaly'' and 
 clarify its meaning.
 
To compute \rf{ki} we note that 
the integral over $\tau $ can  be  performed using that
\be
\int_0^\infty d\te \ { \cosh (j \nu \te ) \ov \cosh ^{s} \kappa  \te}
   = \frac{2^{{s}-2}}{\kappa \ \Gamma(s)}\  \Gamma\left(\frac{s}{2}+\frac{j \nu}{2 \kappa
   }\right)\ \Gamma\left(\frac{s}{2}-\frac{j \nu}{2 \kappa }\right)\ .
\label{formu}
\ee
We then  end up with 
\be 
&& C_{123} = A_1 + A_2   =
 \rc_j   \sql  \k^{-1}  \Big[  2 \k^2 C_1   -  (\k^2  - \nu^2) C_2  \Big] \ , \la{pp}\\
&& 
C_1 =
 {2^{j+1}\over\Gamma(j+2)} \  \Gamma (1+{j\over 2}+{j \nu\over 2\kappa})\ 
 \Gamma(1+{j\over 2}-{j\nu \over 2\kappa})
\ ,\la{piu}\\
&& C_2 = { j+1  \ov j }  {  \k^2  \ov \k^2 - { \nu^2 } }\ C_1 \ .
\label{aqui}
\ee 
If one uses the first procedure (I) setting $\k^2=\nu^2$  in $U_2$ in \rf{ki}
then   the   $C_2$   contribution  is absent.
If one uses the second  procedure (II), i.e. takes $\k^2  - \nu^2\to 0$ 
in the  final result \rf{pp},  then the  $ \k^2  - \nu^2$ factor from  $U_2$ term 
in \rf{ki} cancels against the pole in  $C_2$ in \rf{aqui}. 
This pole comes from the corresponding
$\G$-functions in the integral \rf{formu} and originates from 
the fact that the $\tau $ integral multiplying $U_2$
diverges at $\tau=\infty $ (for $j\nu>0$) if $\kappa\to\nu $.
Thus there is an additional 
``anomalous'' (or ${ 0 \ov 0}$)    contribution. 
As  a result, we find  ($J = \sql \nu$) 
\be 
&& C^{(I)}_{123} = A_1 =  \rc_j \  2 \sql   (\k C_1)_{\k=\nu}
=   \rc_j \    {2^{j+2}\ov j+1}  \ J \ , \la{yuy}\\
&&
 C^{(II)}_{123} = A_1 + A_2 = 
 \big(1 - { j+1 \ov 2  j} \big) \ C^{(I)}_{123}
=  { j-1 \ov 2  j} \ \ C^{(I)}_{123} \la{an} \ . 
\ee 
Using normalization \ci{z,rt,bcfm} of the  CPO vertex given 
 in \rf{koz}, we get for  $\rc_j $ in   \rf{jpl}
\be\la{nor}
\rc_j=2^{-j} \ \bc_j= \frac{(j+1)\sqrt{j}}{2^{j+2} N} ~ .
\ee
This choice
leads to the expected result for the  3-CPO correlator  \ci{Lee} 
 if one uses the first  procedure \ci{z}
\be \la{sta}
 C^{(I)}_{123} =  {1 \ov N}  J \sqrt{j} \approx  { 1 \ov N}  
  \sqrt{  J  (J+j)  j }
  \ . \ee 
  While one  may argue \ci{z} that the  normalization of the  CPO vertex
  is unambiguously fixed by the OPE  argument in \ci{bcfm} 
  here we will formally   consider  the  use of 
 the alternative -- second -- procedure since its result 
 will be   smoothly
 connected  with  that in the case  when the  ``heavy'' state
 is not BPS, i.e.   when one 
  starts  with a  generic string 
 solution and then takes  the string size to zero
 so that  the string state  becomes BPS. 
 This is, in fact,   how one  can understand this second prescription:
 by starting with  a slightly off-BPS expression and taking BPS 
 limit.\foot{We are making an
 implicit assumption that the BPS limit is universal, i.e.   does not depend on
  off-BPS starting point. This appears to be the case as  illustrated on explicit examples
  below.}
 
As follows from \rf{an}, in this second case the  expression in  \rf{sta} 
 is  recovered by changing the 
  normalization  in \rf{nor} to 
  \be\la{nor2}
\rc_j'={2j\over j-1} \rc_j = \frac{j (j+1)  \sqrt{j}}{2^{j+1} (j-1)  N} ~~ .
\ee
We thus  propose that \rf{nor2}
is  the  right  normalization  for computing 
 the semiclassical correlators involving  at least two 
 non-BPS operators  with large 
 charges as well as   chiral primary  vertex operators. 

As will be shown below, it is by adopting this normalization that 
a  general three-point function will  reduce to the standard result \rf{sta} 
in the  limit  when non-BPS operators become CPO.\footnote{An 
alternative possibility, that we shall not  pursue
 here, is that  this  discontinuity 
  may have a physical meaning and may also 
  have a counterpart on the gauge theory side, being, e.g., related to 
  mixing with double-trace operators  (cf. \ci{ruu}).}
  As we shall see below in section 4, the use of this  normalization 
  will also  simplify subleading (in near BPS limit) terms 
  in the 3-point correlators (the coefficient $j-1\ov 2j$
  will  appear as an overall factor  and will cancel against
   similar factor in $\rc_j'$).
  

In  general, the issue of  normalization of the  string vertex operator is subtle;
an invariant meaning has the ratio   of a  3-point function 
of chiral primary  operators \rf{kz}  to their norms, and this depends on a  prescription 
of how  the  3-point correlator is computed. Since this is an example of an extremal 
correlator,  this computation may be subtle  (see below).
Indeed, the  normalization of the chiral primary  operator 
   may be sensitive to the definition of the 
   corresponding supergravity scalars   or the choice  of  basis 
   for the  dual  gauge theory operators \ci{af}.

  To  argue that 
 the  second procedure of computing \rf{ki}  has a natural physical meaning
 let us note that 
  the $\k^2  - \nu^2= 0$ condition  may be interpreted as  the  
 marginality condition \ci{bt1}   for the ``heavy''  vertex operators. 
 In the semiclassical approach  this condition 
 may be  viewed as originating from extremising with respect to 
  the world-sheet metric 
 (cf. also \ci{jan});   in this case it may  be imposed at the end
 of the calculation. 
 Also, since the object  we are computing is essentially 
 a  scattering amplitude, by analogy   with the usual LSZ   relation in
  field theory the mass
 shell conditions  may be implemented in the final expression.  
 
  The above $ \big( { \k^2 - \nu^2  \ov \k^2 - \nu^2}
  \big)_{\k^2 \to \nu^2} $ 
  ambiguity     is reminiscent of a similar ``anomaly''
  in the computation of extremal ($\Delta_1= \Delta_2 + \Delta_3$) 
   3-point correlators 
   of BPS operators in $AdS_5\times S^5$.
   In that case the  supergravity 3-point coupling  vanishes
   as $(\Delta_1-  \Delta_2 - \Delta_3)\to 0$  but  it is  multiplied 
   by the $AdS_5$   integral of the three   Dirichlet  $K$-functions 
   which is proportional 
   to a  product of $\G$-functions  that contains \ci{f1}  a pole 
   $1  \ov \Delta_1-  \Delta_2 - \Delta_3$. 
   One can then  get  the  correct non-zero result  for the extremal 
    correlator   by  a formal analytic continuation
    prescription: first  assume that  
    $\Delta_1=  \Delta_2 + \Delta_3 + {a} $ and then take the limit 
    ${a} \to 0$  in the final expression  \ci{lt}.
    This prescription may be viewed as  formal
    since for the BPS operators dimensions 
     $\Delta_i$ take only integer values\foot{Other  ways to 
    obtain a consistent non-zero result for the  extremal  correlators involve 
    accounting for non-linear redefinition of the supergravity scalars  
    related to mixing with  double-trace operators on the gauge 
    theory side \ci{af}
    or by accounting for  boundary terms in the supergravity action 
    for  the scalar fields \ci{f2}.}
      but it becomes more meaningful 
     in the limit of large (semiclassical)   charges   and, especially, 
     when viewed from the perspective of generic non-BPS string states, i.e., if 
       one   defines  a correlator of the BPS  operators as a 
     limit of a 
     correlator involving non-BPS  string states  and assumes that
      the result should be 
     analytic  in the string-state quantum numbers.

   The 3-point   correlator  of the   chiral primary   operators 
   with $\Delta_i = |j_i|$ is  always  extremal  
   because of  the  charge conservation.
   For example, one may choose  $j_1 = J,\  j_2 = - (J-j) , \ j_3=j$
   and then the case we considered above corresponds to $J \gg j$. 
   Relaxing the $\k = \nu$   condition  is the same as 
   moving away from the point  $\Delta =  |J| $  and is thus   similar to 
    relaxing the condition of 
   extremality  of the correlator.

As we shall see below, in the computations with 
 non-BPS ``heavy'' states there will be no similar  ambiguity.  
  We shall  
 demonstrate  in detail 
 (extending the  observation in \ci{z})
that  it is the second procedure that is   selected 
 if we define the semiclassical correlator \rf{jpl} 
  for  BPS  ``heavy'' state 
 as a limit of   the    correlator where 
 the ``heavy'' state is represented by a semiclassical   string 
 that  admits a point-like  BPS  limit.

\renewcommand{\theequation}{3.\arabic{equation}}
 \setcounter{equation}{0}

\section{BPS limit   of  correlator   with ``heavy'' states as 
 rigid   spinning  strings on $R \times S^5$}

We would like to  demonstrate that the result
 \rf{an} of the second   procedure discussed in the
previous  section is indeed a 
universal outcome of  a similar computation  for a  ``heavy state''
represented by a 
generic semiclassical string 
  in the subsequent    ``small-string'' or  BMN-type   limit. 


As we will  show in this section, this can be done  for any 
solution described by  a rigid rotating string ansatz
   \ci{ART}. 
\footnote{Such strings are ``stationary", i.e. there exists  a rotating frame 
 in 
which they are static.}
Rotating strings can have a diversity of rigid shapes, which are 
determined by the solutions of the Neumann-Rosochatius system. In this 
section we shall consider the case
when the string is point-like in $AdS_5$, i.e. is 
again described by \rf{ga}. Strings extended in $AdS_5$ will 
be discussed in section 4.

The $S^5$ part of the ansatz  will be chosen as 
\be 
&& X_1+iX_2=r_1(\s )  e^{i  w_1  \tau}\ ,
\ \ \ \ \ \ \ \ \ \  w_1 \equiv \nu \ ,\nn \\
&& X_3+iX_4=r_2(\s ) e^{i\a_2(\s )} e^{iw_2 \tau}  \ , \ \ \ \ \ \ \ \ 
X_5+iX_6=r_3(\s ) e^{i\a_3(\s )} e^{i w_3 \tau} \ ,   \la{xxx}
\ee
with   ($i=2,3$; $m_i$ are integer)
\be 
&& \la{ssa}\a_i'={v_i\ov r_i^{2}} \ , \ \ \ \ \ \ \ 
\kappa^2=r'^2_1      + \nu^2 r^2_1+ \sum_{i=2}^3( r'^2_i      + w_i^2 r^2_i  +  
 { v_i^2 \ov r^2_i} )    \ , \\
&& \la{ssb}
\ w_2 v_2+w_3v_3  =0   \ ,\ \ \ \ \qquad v_i \int^{2\pi}_0 { d\s \ov r^2_i(\s)} = 2 \pi m_i 
 \  .
\ee
We have set $\a_3=0 $, that is $v_3=m_3=0$, because we
 are  interested in solutions which smoothly approach the BMN solution
 \be
v_1=v_2=v_3=0\ ,\qquad  r_1=1\ ,\qquad \kappa= \nu\ .\la{bn}
\ee
The equations of motion for $r_s(\s )$ are 
\bea 
&& r''_1   +  (\nu^2 + \L ) r_1  = 0  \ , 
\ \ \ \ \ \ \ \ \ \ \ 
 r''_i   - {v_i^2\over r_i^3}  +  (w^2_i + \L ) r_i  = 0  \ , \ \ \ i=2,3
\la{jk} 
\\
&& \sum_{s=1}^3 r_s^2 =1 \ ,   
\qquad\ \ \ 
 \L=\L(\s) = \sum_{s=1}^3 ({r'_s}^2 +{v_s^2\over r_s^2}  -  w^2_s  r_s^2)\ . 
\la{ppl}
\ee
Let us now consider 
 this  solution (which, in general, carries three $SO(6)$ spins
$J_1,J_2,J_3$)
as representing the ``heavy'' state in  the computation of the 3-point function in
\rf{jpl}.
Using \rf{koz} we obtain  the  following generalization of \rf{ki},\rf{ju} 
(assuming again that  $\tau= -i \tau_e$)
\be
&&C_{123}=   { \sql }  \rc_j  \ \int_{-\infty}^\infty d\tau_e  \int _0^{2\pi}
{d\s\ov 2 \pi} \ 
\left[{r_1(\s ) e^{\nu\tau_e }\over\cosh \kappa \tau_e } \right]^j\   \Big[ U_1 (\tau_e)   +
U_2(\s)\Big] \ , \la{py}\\
&&
U_1 = {2\kappa^2 \over\cosh^{2}\kappa \tau_e }\ , \ \ \ \ \ \ \ \ \ \ 
U_2 = -\kappa^2 - \p^a  X_k \p_a  X_k  = - \k^2  - \L(\s)\ , \la{yp}
\ee
where we used (\ref{ppl}). Since  $\L$  and thus $U_2$ 
do not depend on $\tau_e$, the 
  integral over $\tau_e$ in  \rf{py}   can be performed in the same way as in 
  the previous section (i.e. using  \rf{formu})  and we get 
  the expression  generalizing  \rf{pp}
\be
&& C_{123}=  A_1 + A_2  \ , \la{q} \\
&&  A_1 = 2\sql \kappa\ \rc_j \  C_1\ \int_0^{2\pi} {d\s\ov 2 \pi}  \  [r_1(\s )]^j\ ,\la{qq} \\
&&  A_2 = -\sql \kappa\  \rc_j \  C_2\ \int_0^{2\pi} {d\s\ov 2 \pi}  \  [r_1(\s )]^j     \Big( 1
 + { \L\ov \k^2} \Big)  \  ,  \la{qqq}
 \ee 
where $C_1$  and $C_2$ are the same as in \rf{piu},\rf{aqui}. 

Let us now  expand around the BMN solution \rf{bn}  for which 
$
\L  \to -\nu^2$. 
In this case 
\be
&&\L = -\nu^2 +\epsilon^2 \tilde\L \ , \ \ \ \ \ \ \ \ \   \ep\to 0 \ , 
\label{iu}\\
&& 
r_2= \eps \td r_2 \ ,\qquad r_3= \eps \td r_3\ ,
\qquad r_1=\sqrt{1-\eps^2 \big( \td r_2^2+\td r_3^2\big) } \ ,
\la{riu}\\
&&
v_2=\eps^2 \tilde v_2\ ,\qquad v_3=\eps^2 \tilde v_3\ ,
\ \ \ \ \ \ \ \ 
\kappa^2 =\nu^2+ \eps^2\td \kappa^2\ ,
\la{criu}
\ee
where we assume that the variables  with tilde  are of order 1.
Such 
 solution describes small strings of rigid shape 
 rotating on $S^5$ with  $ J_1 = \sql \nu \gg  J_2,  J_3$.

Using (\ref{iu})-(\ref{criu}) and expanding in powers of $\epsilon $, we find
from \rf{piu},\rf{qq}
\be
&&C_1={2^{j+1}\over j+1} +O(\eps^2) \ ,\la{cac} \\ 
&& A_1= \rc_j ( j+1)^{-1} 2^{j+2} J_1  + O(\epsilon^2)\ , \la{bbv}
 \qquad  \ \ \ \ \ \ 
J_1=  \sqrt{\lambda}\ \nu 
\ee
with the leading term being  the same as in \rf{yuy}.
To compute the  second contribution $A_2$ in \rf{qqq} coming from the $U_2$ term in \rf{yp}
we note that according to \rf{piu},\rf{iu} \rf{criu}
\be
&& \kappa(1+{\Lambda\over\kappa^2})={\td \kappa^2+\td \L\over\nu }\ \eps^2+ O(\eps^4) \
, \la{kio}\\
&& C_2= {2^{j+1}\nu^2\over j \tilde \kappa^2}{1 \ov \epsilon^2} + O(1) \ . \la{klio}
\ee
This  leads to a finite ``anomalous'' contribution
\be
A_2  = -  \rc_j  j^{-1}   J_1  \ \Big(1 + {1 \ov  \tilde \kappa^2}
 \int_0^{2\pi} {d\s\over 2\pi} \    \tilde \L \Big)   + O(\epsilon^2) \ .
\la{aas}
\ee
To  compute the integral over $\s$  we use \rf{jk},\rf{riu}, i.e. 
\be 
&&r_1''+\eps^2\td\L \ r_1=0\ , \ \ \ \ \ \  \ \ \ \  r_1=1+O(\eps^2) \ ,       \la{poy} \\
&&
\td\L = -{r_1''\over r_1 \eps^2}  = 
 (\td r_2 \td r_2' + \td r_3 \td r_3')'+ O(\eps^2) \ . 
\ee
Thus 
 $\td \Lambda $ is a total derivative which  drops out  of 
the integral over $\sigma $ in (\ref{aas}); 
in the limit $\eps\to 0$ we are left with 
\be
A_2  = -   \rc_j\ j^{-1}  2^{j+1}   J_1  + O(\epsilon^2) 
=  - {j+1\over 2j} A_1 +  O(\epsilon^2)   \ ,
\la{aasa}
\ee
which is the direct analog of the expression for 
the  second ``anomalous''  contribution in \rf{an}. 
We  observe that this ``anomalous''   contribution 
is universal for all rigid  string solutions
on $S^5$ which approach the BMN one 
(i.e. it  does not depend on the shape 
of the string $r_i(\s )$ or  parameters $v_i$).

Altogether, we find the same expression for \rf{py}  as in 
\rf{an}, i.e. as in the second procedure of section 3, 
\be
 C_{123} = {j-1\over 2j}  A_1  +  O(\epsilon^2)   \ . \la{ree}
\ee
This  leads us to the standard  3-CPO  result \rf{sta}  
if we  adopt   the modified normalization in \rf{nor2}, 
$\rc_j \to \rc_j'$, i.e. $ C_{123} \to  C'_{123}$, 
\be  
\lim_{\eps\to 0} \ C'_{123}
   =  {1\over N}\ J_1\sqrt{j}  \la{iii} \ , \ \ \ \ \ \ \ \ \ 
   C'_{123}\equiv  C_{123} ({\rc_j \to \rc_j'}) \ . \la{caca}  \ee
We conclude that  to have a smooth BPS limit, 
 the 3-point function
involving ``heavy''  states 
with generic values of the spins ($J_1,J_2 ,J_3 $)
should be computed with the normalization (\ref{nor2})
for the ``light''  chiral primary vertex operator.


\renewcommand{\theequation}{4.\arabic{equation}}
 \setcounter{equation}{0}

\section{Large spin  expansion of  correlators 
  with ``heavy'' states\\ corresponding to spinning   strings in $AdS_5\times S^5$}

Our aim in  this  section  will be to consider several explicit examples 
involving ``heavy'' states represented  by 
 spinning strings in $AdS_5\times S^5$ that admit a point-like (BMN-type)  large $J$
limit  and to  compute  subleading  $1/J$ terms in the 3-point coefficient 
$C_{123}$. We shall see   that these subleading terms have the 
same structure as
expected in perturbative gauge theory, suggesting that their coefficients are
protected, i.e.  should match the gauge theory ones  (see the discussion in the
Introduction). 

The examples we will consider below are: 
 folded spinning string    on $S^5$ (with spins $J_1,J_2$),  folded spinning string 
extended on $AdS_5$ and orbiting $S^5$ (with spins $S,J$), and a 
circular string extended and rotating on 
both  $AdS_5$ and $S^5$ (with spins $S,J_1,J_2,J_3$).

\subsection{Folded  $(J_1,J_2)$ spinning  string in $S^5$}

The solution describing a folded string carrying angular 
momenta $J_1 $ and $J_2$ is a particular example 
of a rigid string solution  on $S^5$ discussed 
in the previous section.
In this case we have \rf{xxx}   with 
 $\a_i=0, \ r_3=0$ and \cite{Frolov:2003xy}
\be
&&  r_1 = \cos \theta  = {\rm dn} ( w_{21} \sigma | q) \ , \ \ \ \ \ 
r_2 = \sin \theta  =  \sqrt{q}\  {\rm sn} ( w_{21} \sigma | q)\ ,\\
&& q\equiv  \sin^2 \theta _0 = \frac{\kappa^2 - w_1^2}{ w_2^2 - w_1^2} \ , \ \ \ \ \ \ \ 
w_{21}^2\equiv 
w_2^2 - w_1^2 =  {2\over \pi}\,  {\rm K} (q)\ .  \la{yty}
\ee
Assuming that the  ``heavy'' state in the 3-point correlator \rf{jpl} 
is represented  by such string carrying two angular momenta $(J_1,J_2)$,  
we find from   \rf{py}\foot{We use that here $\L$ in \rf{ppl} is 
$\L= r_1'^2 + r_2'^2 - w_1^2 r_1^2 - w_2^2 r_2^2$, while 
the conformal gauge constraint in \rf{ssa}  takes the form 
$\kappa^2=r'^2_1 + r_2'^2 + w_1^2 r_1^2 + w_2^2 r_2^2$.} 
\bea
C'_{123} =  2\sql\  \rc'_j\,  \int_{-\infty}^\infty d\tau_e
  \int^{2\pi}_0 {d \sigma\ov 2\pi}
\left[ \frac{ \cos \theta (\s)   e^{ w_1 \tau_e}   
   }{\cosh \kappa \tau_e }  \right] ^j
\left[ {\kappa^2\over \cosh^2 \kappa \tau_e } - \kappa^2 -  
w_{21}^2 \cos^2 \theta (\s)  + w_2^2 \right] . 
\label{tresJ}
\eea
This  example was previously considered in \ci{z}.  Here we  put prime on $C_{123}$ 
since we will use 
 the  modified normalization $\rc'_j$ of the ``light''
 chiral primary  operator in \rf{nor2} to get  a smooth BMN-type limit 
 when $ J_1 \gg J_2$ as discussed in the previous section.
We  will be interested in the explicit form of the subleading corrections 
to  the leading CPO term \rf{sta}.  

According to \rf{q}--\rf{qqq} we get 
 $C'_{123} =A_1+A_2$ with
\bea
A_1 &=& 8\sql\ \kappa  \ \rc'_j\ \ C_1\ \int_0^{\pi\over 2} { d\s\ov 2 \pi}  \ \big[\cos \theta (\s)
\big]^j  \ , 
\la{juo}\\
A_2 &=& -{8\sql\ \kappa^{-1}}\ \rc'_j\ \ C_2\ \int_0^{\pi\over 2}{ d\s\ov 2 \pi} 
 \  \big[\cos \theta (\s)
\big]^j \left[ 
 \kappa^2  +  w_{21}^2 \cos^2 \theta (\s)  - w_2^2   \right]\ , \la{qyi}
\eea
where  $C_1$ and $C_2$ as in  (\ref{piu}),(\ref{aqui})  with $\nu=w_1$.
We took into account that for folded string the   $\s$ integration region 
is naturally split into 4 equivalent parts.


Defining  
\be  \J_i = { J_i \ov \sql}\  , \  
 \ \ \ \ \   \ax = {\J_2\ov \J }\  , 
\    \ \ \ \ \    \calJ= \J_1 + \J_2\  , \la{kot}\ee
we have $\ax    \ll 1 $  in the near-BMN limit when  $\J_1 \gg \J_2$ .
We shall first 
expand in powers of $1\ov \J_i$  and then expand each
 coefficient of this expansion in powers of $\ax $.
In doing so one finds
\cite{Frolov:2003xy}
\be
&& w_1= \calJ   -  \frac{\ax}{2 \calJ  } (1 +   \ax + ...)+...\ ,\qquad \ \ 
w_2= \calJ   +  \frac{1}{2 \calJ  } (1  + {\ax^2\over 8}+...  )+... \ , \\
&&
\kappa= \calJ  +   \frac{\ax}{2 \calJ  }  (1 +  \frac{\ax}{2} + ...)+...\ ,\qquad \ \ 
q = 2 \ax ( 1 -  \frac{\ax}{4} -{\ax^2\over 8}+ ...) \la{qw} \ , \\
&&
\theta (\sigma) =  \sqrt{2 } \Big( \ax^{1/2}      \sin (\sigma ) + 
{1 \ov 24} \ax^{3/2} \left[ 6   \sin (\sigma ) +
  \sin (3 \sigma) \right] + ...\Big) \ . \la{ppt}
\ee
Then the  second factor in the   integrand of $A_2$ in \rf{qyi} has the following
 expansion  in small  $\ax= {\J_2 \ov \J} $
\be 
  \kappa^2  +  w_{21}^2 \cos^2 \theta (\s)  - w_2^2  
 =\Big[ 2 \cos ^2 \s       -\frac{1}{4 \J^2} \big( 1 +  2 \sin ^2 \s\big) \Big]\ \ax\ 
   +O\left(\ax^2\right) \ .
\ee
As in the general case of a rigid spinning string \rf{kio}, 
it     vanishes in the point-like limit ($\J_2\to 0$).
Using \rf{aqui}, we find that   as in \rf{cac},\rf{klio} 
\be
C_1= {2^{j+1}\over j+1}+O(\ax )\ ,\qquad \ \ \ C_2= {2^{j}\over j}\ {{\cal J}^2\over\ax}+O(1 )
\ . \ee
Thus we  obtain   the same expressions  \rf{aasa},\rf{ree},\rf{iii} 
as the leading term in the   $\J_2\to 0$ or  $\ax\to 0$ 
limit.


Including corrections in $\J_2 \ov \J$,  we observe 
that the   ratio ${ j-1\over 2j}$ factors out  also from the subleading terms 
and is thus cancelled against a similar factor  in the normalization coefficient 
$\rc_j'$ in \rf{nor2}.  We  thus end up with
\foot{This is, of course, the leading semiclassical approximation, 
i.e. we ignore  string $\a' \sim {1 \ov \sql}$ corrections 
that lead  to terms suppressed by extra powers of ${1 \ov J}= {1 \ov \sql} { 1\ov \J}$.} 
%
\be
&&C_{123}' =
{1\over N}\  \sql\ \J \sqrt{j} \ 
\bigg[ 1 - \frac{\ax }{4} \, (1+3j)  +  O(\ax^2) \nn \\
&&\ \ \ \ \ \ \  \ \ \ \ \ \ \ \ \ \qquad  \ \ \ \ \ \ \ +
 \ \frac{\ax}{8 \J^2}\  \Big(   5+3j-4j\big( \psi (j)+\gamma \big) + O(\ax)
 \Big) +O\big({\ax\over \J^{4}}\big) \bigg] \ ,  
\ee
where $\psi(x)=\Gamma'(x)/\Gamma(x)$ and $\gamma $ is the Euler's  constant. 
Equivalently, in terms of the  spins  $J_1,\ J_2$, and $\l$  this reads, in the limit
when $J_1 \gg J_2$, 
%
\be
&& C_{123}'= {1\over N}\ J_1\sqrt{j} \ 
\bigg[ 1 - \frac{3J_2 }{4 J_1} (j-1) 
-\lambda \frac{J_2}{8 J_1^3 } \ \Big(4j \big(\psi (j)+\gamma -
 {\textstyle{3 \ov 4}}\big)-5 \Big) \nn\\
 && \ \ \qquad \qquad \qquad  \qquad   +\  O\big({J_2^2\over J_1^2}\big)
 + O\big( { \l J^2_2 \ov J_1^4}\big)  
 +   O\big( { \l^2 J_2 \ov J_1^5}\big)
\bigg]  \ .    \la{kut}
\ee
Observing that the charges  of the three operators
entering the correlator  may  be assumed to be 
$(J_1,J_2), (-J_1-j, -J_2), (j, 0)$ we conclude that
for $J_1 \gg J_2 \gg j$  the leading term is 
the  standard  protected 3-CPO correlator. 

The two explicitly  written subleading terms in \rf{kut} look like  the  
 tree-level and the one-loop 
corrections  on gauge theory side.
\foot{For example, the  1-loop  correlator of 3 near-BMN operators with
large charges $-J,\  r J$ and $ (1-r) J$  has the following form  \ci{okk}:
\ 
$C_{123}= {  1\ov N} \sqrt{J^3 r(1-r)}   \Big(a_1  + a_2 { \l \ov J^2} + ... \Big)$
where $a_1$  and $a_2$  depend on detailed  structure of the operators 
(number of impurities, etc.).} 
We   conjecture  that their coefficients are also  
  not  renormalized, i.e. they 
should  be the same in the perturbative string ($\l \gg 1$) and in the 
perturbative 
gauge ($\sql \ll 1$) theories.\foot{It is likely that the same
non-renormalization should apply also to the coefficient of the next 
term we did  not write explicitly  which  is proportional to 
$\l^2$.}
It thus remains  to compare \rf{kut}   in detail 
with the  one-loop $\cal N$=4 SYM  correlator  of the corresponding
$SU(2)$ sector operators in the limit $J_1 \gg J_2 \gg j$.



\subsection{ Folded  $(S,J)$ spinning string in $AdS_5\times S^1$}

Let us now repeat  the same computation in the case when the ``heavy'' state is
 represented by the folded spinning string in $AdS_3$ part of $AdS_5$ orbiting big
 circle of $S^5$  \ci{gkp,ft1}.  In the limit of large spin  this example 
  was already discussed in \ci{rt}. 
For generic values of  spins $(S,J)$   the solution written in global $AdS_5$ coordinates is 
\be
&&\r=\rho(\s), \quad t= \ka \tau, \quad \phi= w \tau, \quad \varphi= \nu \tau,\la{ewd} \ \ \ 
{\rho'}^2 = \ka^2 \ch^2 \rho  - w^2 \sh^2 \rho   - \nu^2, \\
&&\sinh\rho = \epsilon\ {\rm sn}
(\sqrt{\kappa^2-\nu^2}\ \epsilon^{-1}\ \sigma,-\epsilon^2)
\ , \ \ \ \ 
\epsilon^2\equiv {\kappa^2-\nu^2\over w^2-\kappa^2}, \nn \ \ \ 
\sqrt{\kappa^2-\nu^2}=\epsilon \ {}_2F_1({1\over 2},{1\over 2},1;-\epsilon^2)
\ee
$\rho(\s )$ varies from 0 up to its  maximal  value given by
$
\coth^2\rho_{\rm max}={w^2-\nu^2\over \kappa^2-\nu^2}=1+
{1\over \epsilon^2}
$
and the angular momenta are
\be\la{sj}
{ S}= \sql \S ={\sql \ w\epsilon^2\over 2\sqrt{\kappa^2-\nu^2}}
\ {}_2F_1({1\over 2},{3\over 2},2;-\epsilon^2)\ ,\ \ \qquad J=\sqrt{\lambda }\ \J,\ \ \  \ \J=\nu  .
\ee
Written in the  Poincar\' e coordinates  and continued 
to the euclidean time the $AdS_5$ part of the solution is \ci{bt1,rt} (cf. \rf{ga})
\be
&&Y_4=0,  \qquad Y_5= {1\over z} = \ch \rho\   \cosh\ka \tau_e , 
 \qquad Y_{0e}= {x_{0e}\over z} = \ch \rho\  \sinh \ka \tau_e \ , \la{ggh}\\
&&
z= \frac{1}{ \ch\rho\  \cosh\ka \tau_e }\ ,\qquad 
x_{0e}= \tanh\ka \tau_e  , \qquad 
x_1+ix_2= \frac{\th\rho }{\cosh\ka \tau_e } \ e^{w \tau_e } \la{ktr} \ . 
\ee
For  the 3-point correlator of the states  with charges 
$(S,J), (-S, -J-j)$ and $(0,j)$   where    $J \gg j$
we find from \rf{jpl} the following analog of the expression \rf{tresJ} 
\bea
C'_{123} =
 8\sql \rc'_j   \int_{-\infty}^\infty  d\tau_e  \int^{{\pi\ov 2}}_0 {d \sigma\ov 2 \pi}
\Big[\frac{    e^{\nu \tau_e} }{ \cosh \rho(\s)\ \cosh \kappa \tau_e }\Big]^j  
\bigg[   \frac{\kappa^2 }{\cosh^2 \kappa \tau_e} -(w^2-\nu^2) \tanh^2 \rho(\s)    \bigg]
\label{awa}
\eea
Let us consider the short string limit when  $S\ll J$, 
i.e. $\ep \to0$  and   $\rho_{\rm max}=\epsilon+O(\epsilon^3)$ so that 
$\epsilon$ measures the size of the string.
 We will expand \rf{awa}
  in powers of $\epsilon$, i.e.  approach the 
 BMN limit from the $\epsilon\not=0$ region. 
 As  in the previous sections, we will then reproduce the 3-CPO  result \rf{iii}
 if we use normalisation constant in \rf{nor2}.

Explicitly, expanding in $\ep$ we get 
\be
&&\rho= \epsilon   \sin  \sigma 
 - { 1 \ov 12} \epsilon^3  ( 3   \sin  \sigma - \sin^3  \sigma) 
  +O(\epsilon ^5)
\ , \ \ \  \ \       \epsilon^2={2\S\over \J}+ {\S^2\over 2\J^2}+...         \\ 
&&\kappa^2  =\nu^2+\epsilon^2-{\epsilon^4\over 2} + ...\ ,
\qquad 
w^2 = 1+\nu^2+{\epsilon^2\over 2}+...
 \ . \la{k}
\ee
Computing the integral over  $\tau_e $ we get
$C'_{123} =A_1+A_2$ with  (cf. \rf{qq},\rf{qqq} and \rf{juo},\rf{qyi}) 
\bea
A_1 &=& 8\sql\ \k \ \rc'_j\  C_1\ 
\int^{{\pi\ov 2}}_0 {d \sigma\ov 2 \pi}  \  {1 \over \cosh^j  \rho(\s)} \ , \la{w}
\nn\\
A_2 &=& -8\sql\ \rc'_j\  C_2\ \k^{-1} (w^2-\nu^2)
 \int^{{\pi\ov 2}}_0 {d \sigma\ov 2 \pi}  \ {\sinh^2
 \rho(\s) \over\cosh^{j+2}  \rho(\s)} \la{ww}  \ , 
\eea
where  $C_1$ and $C_2$ are again given in (\ref{piu}),(\ref{aqui}).
Expanding in powers of $\epsilon $, we 
find that,   again, $C'_{123}\sim  { j-1\over 2j} \rc'_j  $ 
with $ { j-1\over 2j}$ factorising  out also  of the  subleading terms.
As a result, we obtain 
%
\be
C_{123}'= {1\over N}\ J\sqrt{j} \ \bigg[1 -\frac{ \epsilon ^2 \left(j-1\right)}{8 }
-\frac{\epsilon ^2}{4\J^2}
 \ \Big( j \psi(j)+j\gamma -1\Big)+O(\epsilon^4)
+O({\epsilon^4\over\J^2})  
\bigg]  , 
\la{afd}
\ee
or, equivalently,   
%
\be
&&C_{123}'= {1\over N}\ J\sqrt{j} \ \bigg[1 -  {S\over 4J} (j-1) \ 
 -  \lambda {S\over 2 J^3} 
   \Big( j \big( \psi(j)+\gamma\big)-1 \Big) \nn \\
&& \qquad\qquad \qquad\qquad    \  +\   O\big({S^2\over J^2})
+ O\big({ \l S^2 \ov J^4}\big)  + O\big({ \l^2 S^2 \ov J^6}\big)   
 \bigg] \ . \la{ete}
\ee
This is very similar in structure to the expression \rf{kut} in the 
case of the $(J_1,J_2)$ folded string. Note that 
there is no other $O({1\ov \J^n})$ correction to the $O(\epsilon^2)
 $ term in (\ref{afd}), i.e. the next subleading term 
comes from $O({\eps^4\ov \J^4})$, i.e.
 is  $O({ \l^2 S^2 \ov J^6})$.

 Once again, we conjecture  that the leading terms in \rf{ete} should 
 match the leading   terms in the 3-point correlator
 of the one-loop  $\cal N$=4 SYM  correlator  of  the corresponding
$SL(2)$ sector operators with charges $(S,J), (-S, -J-j)$ and $(0,j)$
in the limit   $J \gg S, \  J \gg   j$.


\subsection{Circular spinning  string in $AdS_5\times S^5$}

Another example involves a rigid 
circular string  with spin $S$ in
one plane in $AdS_5$ and  spins  $J_1, J_2, J_3$ in the
 three orthogonal planes of $S^5$.
 It  allows us to provide  another  
 test of the universality of the ``anomaly''\rf{ree} 
 and thus  of the modified normalization  \rf{nor2}
 in the case  when the string is extended 
in both $AdS_5$ and $S^5$ directions.\foot{The corresponding 3-point
 function was  recently computed also in the first paper in ref. \ci{ry}
but the fact that taking the BMN limit in the final   expression 
leads to an additional 
  ``anomalous'' contribution and thus requires to use the normalization \rf{nor2} 
   was not noticed there.}

This is a  solution of the 
general 
$AdS_5 \times S^5$ rigid string ansatz  (cf. \rf{xxx}) 
 in which the radii  $r_i$ are  constant,
i.e. it represents a circular string which winds and rigidly
 rotates  in several planes \cite{ART} 
\be \la{rell} &&
Y_5 + i Y_0 = b_0  e^{i \k \tau  } \ , \ \ \ \ \ \ 
Y_1+ iY_2 = b_1 e^{ i k \s} e^{i \omega_1\tau }\ , \ \ \ \ \ \ \
Y_3 + i Y_4  = 0\  , 
\\   &&
X_1+iX_2=a_1   e^{iw_1\tau} , 
\quad
 X_3+iX_4=a_2   e^{ i m_2 \s} e^{iw_2\tau} , 
\quad
X_5+iX_6=a_3   e^{ i m_3 \s} e^{iw_3\tau} 
\la{embi}
\\
 && w_1 = \nu \ ,\ \ \ \ \ \ \ a_1^2+a_2^2+a_3^2=1\ ,\qquad \qquad b_0^2-b_1^2=1\ , 
\la{ares}\\
&&  k \S + m_2 \J_2 + m_3 \J_3 =0 \ , \ \ \ \ \ \  S= {  \sql \S} \ , \ \   \ 
J_i = { \sql  \J_i } \ . 
\ee
Here $k$ and $m_i$ are integer winding numbers. 
We have set $m_1=0$ to have a smooth BMN limit  when the string becomes  point-like.
For simplicity we ignore  the possibility of the second spin component in $AdS_5$. 

 Computing (\ref{jpl}) in this case  we find (cf.\rf{ki} and \rf{py}  where now
$\Lambda= -\nu^2$) 
\be
C'_{123} = \sql \  \rc'_j  \  \Big({a_3\over b_0}\Big)^j \int_{-\infty}^\infty d\tau_e\ {e^{j\nu\tau_e }\over 
\cosh^j \kappa\tau_e }\ \Big({2\kappa^2\over \cosh^2 \kappa\tau_e }
-\kappa^2+\nu^2\Big)\ .
\ee
As in the previous sections, we get  $C'_{123}= A_1+A_2$, with 
\be
  A_1 = 2 \sql\  \kappa\ \rc'_j  \ \Big({a_3\over b_0}\Big)^j \  C_1\ ,\  \ \ \ \ \ 
\ \ \ A_2 =  -  \sql\   \kappa^{-1}\ \rc'_j  \Big({a_3\over b_0}\Big)^j\   C_2\ ( \kappa^2- \nu^2)
\label{caf}
\eea 
where  $C_1,C_2$  are as  in (\ref{piu}),(\ref{aqui}). Note that 
this is formally the same as  (\ref{pp}) except for the factor $     \big({a_3\over b_0}\big)^j     $.
In the limit when $J_1\gg J_{2}, J_3, S$, we have that $\kappa\to\nu $. 
Then,  once again,  the pole in $C_2$ (\ref{aqui})
cancels against the zero of the numerator 
in (\ref{caf}), producing 
an extra finite contribution 
as compared to the strict BMN limit
 where $\kappa= \nu $ is imposed before the integration over  $\tau_e $.

Assuming that  $\J_i, \S \ \gg \ m_i, k$ 
we get the following expansions 
(see \ci{ART} for details):
\bea
\nu &=& \J -{1\over 2\J^2}\left(\J_2 m_2^2+\J_3 m_3^2\right)
+ {1\over 8\J^5}\left[3\J(\J_2m_2^4+\J_3m_3^4)-4(\J_2 m_2^2+\J_3 m_3^2)^2\right]+O\big( \J^{-5}\big),
\nn\\
\kappa &=& \J +{1\over 2\J^2}\left(\J_2 m_2^2+\J_3 m_3^2+2\S  k^2 \right)
\nn\\
& & \  - {1\over 8\J^5}\left[\J(\J_2m_2^4+\J_3m_3^4+4\S k^2)+4k^2 \S\big( 2\J_2 m_2^2+2\J_3
m_3^2+3\S k^2\big)
\right]+O\big(\J^{-5}\big) ,\la{ops}
\eea
where $\J\equiv \J_1+\J_2+\J_3= {  J \ov \sql}$. 
Then we find from \rf{caf} that again  the overall   ${ j-1 \ov 2 j}$ factor cancels against  the
normalization factor $\rc_j'$ in \rf{nor2}, i.e. 
%
\be
&&C'_{123}= \ {1\over N}\ J\sqrt{j}\  \Big({J_1\over J+S}\Big)^{j/ 2}\ \Big(  1 +  {\lambda\over
2J^3}  P_1 + O\big({\lambda^2\over J^4}\big) \Big) \ , \la{paa}\\
&& P_1=    k^2S -  \Big( J_2m_2^2+J_3m_3^2+k^2 S \Big)\bigg( j \big(\psi(j) + \gamma-1 \big) -1
+  {j \, J\over 2(J+S)}\bigg)\ . \la{klp}
\ee
The factor $ \big({J_1\over J+S}\big)^{j/2} $ originates
 from $\big({a_3\ov b_0}\big)^j$. 
 Note that this expansion applies for general
  (large) values of  $J_1, J_{2}, J_3, S$ , i.e. 
  so far  we did not 
  assume a near-BMN limit.

 In the near-BMN or  point-like string 
  limit $J_1\gg J_{2}, J_3, S$  (when  
   $a_3, b_0\to 1$,  $\Big({J_1\over J+S}\Big)^{j/ 2}\to 1$) the 
    expression \rf{paa}  for 
  $ C'_{123}$ takes the form similar to the one in  \rf{kut} and  \rf{ete} 
  with  the  leading term in the ``one-loop'' coefficient $P_1$ simplifying to 
 \be  P^{(1)}_1=    k^2S -  \Big( J_2m_2^2+J_3m_3^2+k^2 S \Big)\Big( j
  \big(\psi(j) + \gamma- {\textstyle{1 \ov 2} }
  \big) -1\Big) \ . \la{ktp}
\ee
As in the previous examples \rf{kut},\rf{ete} the leading term in \rf{paa} is then 
 the  3-CPO  correlator  and subleading terms 
   look  like 
tree-level and one-loop   gauge-theory 
terms expanded in the  limit $J_1\gg J_{2}, J_3, S\gg m_i, k$. 
Again, it  would be  very interesting to compare these 
expressions to the  gauge theory ones 
for the correlator of the corresponding operators with charges
$(S; J_1,J_2,J_3), \ (-S; -J_1-j ,-J_2,-J_3)$ and $ \ (0; j, 0, 0)$. 


\

\section*{Acknowledgements }

We are  grateful to  E. Buchbinder  and R. Roiban 
 for  collaboration on related problems.
 We also acknowledge R. Roiban  and K. Zarembo 
 for  helpful remarks on the draft. 
 We also thank F. Alday,   
   N. Gromov, G. Korchemsky  and P. Vieira
   for   discussions. 
J.R. acknowledges support by MCYT Research
Grant No.  FPA 2007-66665 and Generalitat de Catalunya under project 2009SGR502.

\newpage


\end{document}